# Development of the FASPAX IC for a high burst rate X-ray imager with very high dynamic range ($10^5$) capability in small pixels

Tom Zimmerman[1], Sachin Junnarkar[2]

[1]*Fermi National Accelerator Laboratory, PO Box 500, Batavia, IL 60510, USA, tzimmer@fnal.gov*
[2]*Field Viewers, 50 E. Old Mill Rd., Lake Forest, IL 60061, USA, junnarkar@fieldviewers.com*

**Abstract**

The technique of current splitting is presented as part of an integrated circuit development for an X-ray imager. This method enables integration of charge signals of unprecedented magnitude in small pixels, achieving a dynamic range of $10^5$. Results from two front end prototypes are given and a final optimized design is proposed.

## 1. Introduction

The FASPAX (Fermi-Argonne Silicon Pixel Array X-ray detector) integrated circuit is being developed as part of a proposal to build a wafer-scale X-ray camera for the APS upgrade at Argonne National Laboratory. This camera will have small pixels (100 μm x 100 μm) with an unprecedented dynamic range of $10^5$ (single photon to 100,000 photons, with a photon energy equal to 8 keV). Fig. 1 is a block diagram of one pixel. The front end of each pixel must acquire and integrate input signal charges at a 13 MHz burst rate. A multi-range scheme is required for such a large dynamic range. For each signal of the burst, the back end must decide on the fly which of the multiple ranges is appropriate based on the signal amplitude, and store that result in an analog pipeline (capacitor storage array). After the signal burst is complete, the chip must then read out all pipeline cells of all pixels in 10 ms or less.

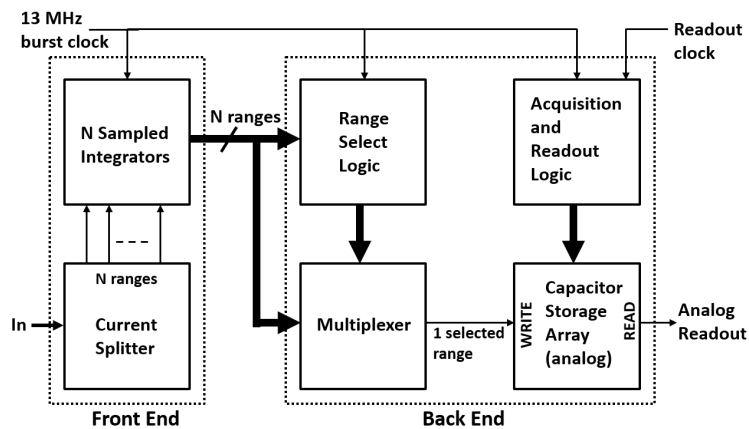

Fig. 1. Block diagram of one FASPAX pixel.

One of the biggest challenges in meeting the dynamic range specification is the integration of such a large signal amplitude (approximately 30 pC for the proposed detector) in a pixel area of only $10^4$ square microns. This capability is approximately an order of magnitude beyond any previous achievement for a pixel of similar dimensions [1–4]. This paper focuses on the design of the front-end and introduces the current splitting technique for pixels, in which the achievable dynamic range is significantly enhanced by accurately fractionating the input charge signal.

## 2. Front End Design Approach

### 2.1 Existing Technique and Limitations

The prevalent existing technique for integrating a signal charge of wide dynamic range in a pixel is with an adaptive gain active integrator, as shown in Fig. 2. Multiple integration capacitors can be switched into the feedback loop of an amplifier depending on the signal amplitude. This forms a closed-loop active integrator with several different sensitivities. Typically, the value of $C_1$ is relatively small, giving high sensitivity on the first range. If the signal is larger than can be absorbed on $C_1$, then additional feedback capacitors of progressively larger value are automatically switched in, resulting in a multi-range active integrator with lower sensitivities for larger signals.

The maximum charge that can be integrated with this technique depends on the available voltage swing (typically around 1 volt) and the maximum value of integrator capacitance. If the largest charge to be integrated is 30 pC, a total integrator capacitance of approximately 30 pF would be required. A linear capacitor of this magnitude cannot be physically realized in the FASPAX pixel area. In addition, in this closed-loop configuration, a fast response image current with a peak of up to several mA per pixel would need to be sourced from the power supply to match the input current. This can be done, but it requires area and power. A completely different approach is required for the FASPAX pixel.

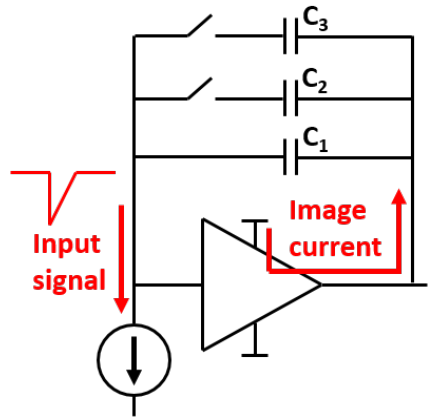

Fig. 2. Adaptive gain active integrator.

### 2.2 Current Splitting: A New Approach for Higher Dynamic Range in Pixels

The technique of current splitting has been used by QIE (Charge Integrator and Encoder) chips in High Energy Physics calorimeters successfully for many years [5]. In the FASPAX front end, this technique is modified for use in small pixels. As shown in Fig. 3, the heart of a current splitter consists of some number of identical bipolar NPN transistors whose base-emitter junctions are all connected in parallel, and whose collectors are ganged as desired to form multiple scaled ranges. If the NPN transistors are perfectly matched, the current output of any given range will be a constant fraction of the input current, based on the ratio of the number of collectors on that range to the total number of collectors. The same would be true if the NPNs were replaced with perfectly matched MOS devices. However, in the presence of inevitable random device mismatches, the actual split ratios will be somewhat different than calculated. In a MOS device splitter with mismatch, those split ratios would not remain constant as a function of the current due to the square-law behavior of the devices. But with a bipolar NPN splitter, although the split ratios are indeed affected by mismatch, those ratios remain constant over a large dynamic range due to the exponential nature of the response of collector current to base-emitter voltage. A current splitter must therefore employ bipolar input devices to offer constant-ratio splitting.

The radiation hardness of bipolar NPN transistors is not well known for all available SiGe BiCMOS processes. For this reason, all bipolar NPN transistors in the physical layout of the pixel will be located underneath the relatively large bump-bond pad. The delivered radiation dose is expected to be attenuated by a factor of several hundred due to this bond pad and the solder bump attached to it. Determining the radiation tolerance of this configuration is not within the scope of this work and will have to be measured at a later time.

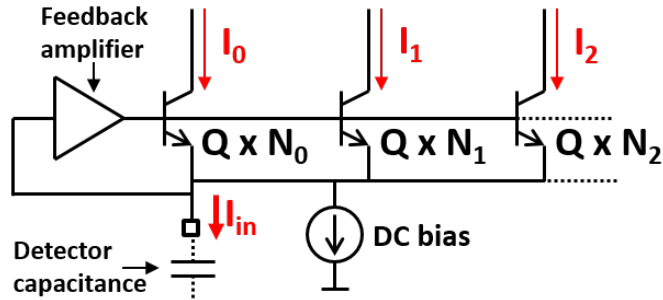

Fig. 3.  Current splitter configuration.

Fig. 3 also shows other elements that are required in the current splitter. The splitter structure is essentially a common-base amplifier with multiple outputs. Some DC bias current must be supplied at the input (emitters) to establish an operating point for the NPN transistors. An external detector capacitance is present at the input. If the NPN bases were simply grounded (forming an open-loop common base structure), the required bias current for low impedance and adequate frequency response at the input would be unreasonably large. Therefore, a feedback amplifier must be incorporated as shown in Fig. 3 to achieve low input impedance with relatively small DC bias current. Typically, the number of splitter NPN transistors ganged on Range 0 (the most sensitive range) is significantly larger than for the subsequent (higher) ranges. Therefore, most of the input current flows to the lowest range, and only a small fraction of the input current flows to the less sensitive ranges. In this way, most of the range scaling can be done *not* by scaling up the value of integrator capacitors on the higher ranges, but by scaling down the input current. This allows the integrator capacitor values on all ranges to remain relatively small.

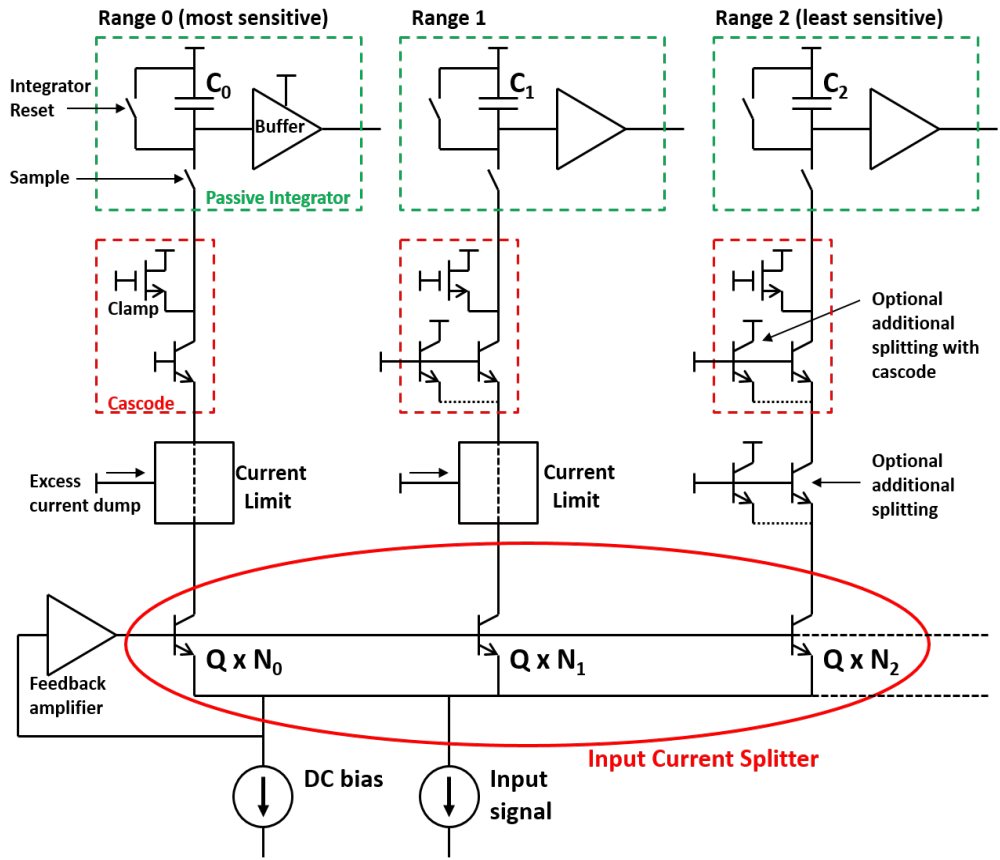

Fig. 4.  Front-end generalized design approach.

Fig. 4 shows the generalized design approach for implementing a pixel front end using the current splitting technique. In addition to the input current splitter, several other important elements are required: passive integrators, current limiting/dumping, cascodes, and additional splitting. Each of these will be addressed.

Passive (open-loop) integrators are much preferred over active integrators in this application because they are very simple and have low power and area requirements. Each passive integrator includes just a small value capacitor, simple Reset and Sample switches, and a voltage buffer which is required to drive the back-end circuitry of the pixel. Each passive integrator is fed by a small cascode transistor, which serves to isolate the integrator capacitance from the significant capacitance of multiple NPN splitter collectors. A clamp transistor on the collector of each cascode transistor keeps it from saturating. On at least the most sensitive range (Range 0), and perhaps on some subsequent ranges, a current limit circuit must be inserted before the cascode transistor. Range 0 is the range of interest for the smallest signals, so the current limit must pass small signals unadulterated. However, for a large input signal (when the higher ranges are of interest), a large current will flow through the Range 0 collectors that would exceed the rating or capacity of the Range 0 cascode transistor. Most of this current is therefore diverted ("dumped") by the current limit circuit to an off-chip virtual ground node. In this way, a large fraction of a small input signal is integrated on Range 0, whereas a large fraction of a large input signal is dumped by Range 0 while a small fraction of that large signal is integrated on a higher (less sensitive) range. A virtual image current of the signal never needs to be sourced by the power supply, as would be required with the adaptive gain approach with an active integrator. Here all the work of integration is conveniently done by the signal itself. Each front-end range has its own integrator, so that all ranges simultaneously integrate different fractions of the input signal.

On the higher ranges, additional splitting can be incorporated before the cascode, and as part of the cascode. This allows realization of large-scale factors between ranges without using large numbers of NPN transistors in the input current splitter. These additional splitters typically do not require incorporation of feedback amplifiers (as does the input splitter) because their inputs are not loaded with a significant capacitance.

Fig. 5 is a timing diagram showing the time structure of the front-end control signals during the 13 MHz burst (an input signal arrives every 75 ns). All integrators are initially reset for 10 ns, followed by a 50 ns integration period. The bandwidth of the splitter, current limit, and cascode combination, which is directly dependent on the DC bias current, must be high enough that the signal is fully integrated in 50 ns. At the end of the integration period, the Sample switch is opened, and another 15 ns is allowed for the integrator voltage buffers to settle, and for the logic in the back end to decide which range is of interest and then multiplex it to the storage array.

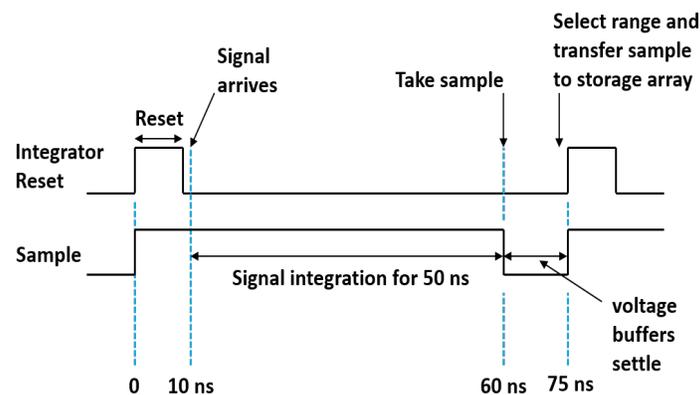

Fig. 5. Front-end control signal timing.

The advantages of the current splitting technique for pixels are significant. Since the multi-range scaling is done by fractionating the signal as opposed to multiplying the integration capacitance, all integrator capacitors can be small, taking only a very small fraction of the total pixel area. Also, active sourcing of a signal image charge is not required. The biggest challenge of the current splitting technique is in managing the effects of the required DC bias current. This current adds to the signal, producing different "offsets" on different ranges. It also contributes shot noise. These effects are addressed in the following sections.

# 3. Front End Prototypes Designs and Results

## 3.1 First FASPAX Front End Prototype Design

Fig. 6 is a simplified schematic of the first FASPAX front end prototype. The main objective of this prototype was not to arrive at a final optimized design, but to demonstrate that the current splitting technique is viable and effective in a small pixel. The HP BiCMOS8HP 130 nm SiGe process is used for all prototyping.

For simplicity, three ranges were chosen (High Gain – HG, Medium Gain – MG, and Low Gain – LG), where the MG and LG ranges are each scaled by a factor of 32 from their neighbors. All the integrators have a 500 mV linear range and are identical, and all of the range scaling is done with current splitting. The input splitter has 32 NPNs on the first range and one NPN on the second, achieving the HG:MG scaling factor of 32. The cascode of the LG range incorporates a second current splitter, giving the MG:LG scaling factor of 32.

A range map of this configuration is shown in the lower right of Fig. 6. One photon produces about 0.3 fC (200 e−) with the planned detector. With a 105 fF integrator capacitance (including parasitics) and 500 mV integrator range, the HG range integrates 1 to 167 photons, the MG range 167 to 5330 photons, and the LG range 5330 to 170K photons, for a dynamic range of more than $10^5$. The integrated charge will ultimately be converted by an external ADC. The number of bits required for that ADC is set by the condition that the ADC binning noise always be smaller than the electronic circuit noise for the smallest signal, and that it is smaller than the inherent Poisson noise of the signal at all other signal levels. The hardest place to meet the latter condition is always at the bottom of a given range, where the Poisson noise is the smallest for that range and ADC bin width. In this system, a 10-bit ADC would be required to meet this condition at the bottom of the LG range. Since the function of this first prototype was to serve as a proof of principle for the current splitting technique, no consideration was given to the circuit noise for the smallest signals.

Current limiting is performed by adding a series resistance in the first splitter output at Node A and clamping that node with the emitter of a current dump transistor (and likewise in the second splitter output). For small currents, the dump transistor is off, and at some larger current amplitude it turns on and shunts away most of any additional current from the cascode.

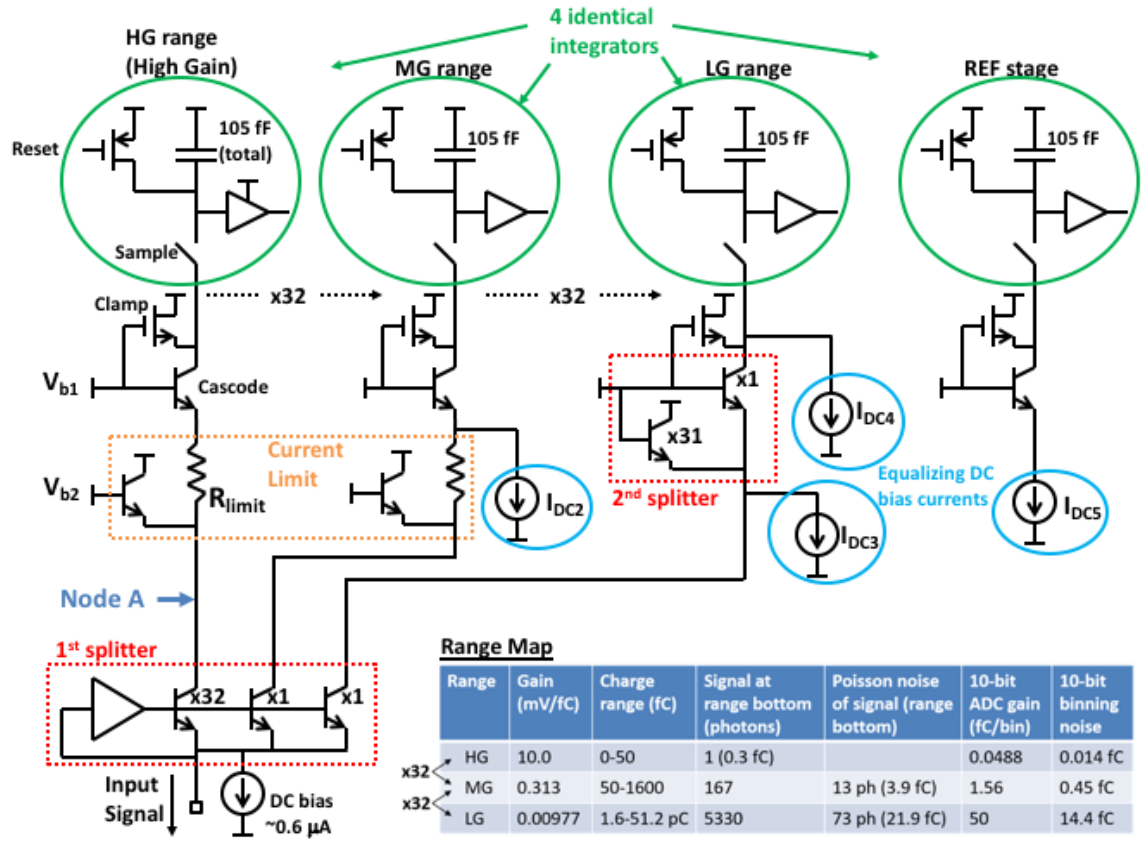

Fig. 6. Simplified schematic of the first front-end prototype

In this circuit, the small signal frequency response is limited by Node A. The 32 collectors and other sources of capacitance on Node A, in conjunction with the cascode small-signal emitter impedance, form the dominant pole of the circuit. The cascode emitter resistance is directly proportional to its DC bias current, and a DC input bias of approximately 0.6 mA is required to give a frequency response at Node A that is adequate to completely integrate the signal within 50 ns. This significant bias current is integrated along with the signal on range HG, producing a voltage offset of about 300 mV. This adds to the signal and is quite vulnerable to variables like integration clock jitter, bias current drift, etc. One way to reduce this vulnerability is to add a Reference (REF) integrator (as is done in QIE chip designs), which is simply another integrator that receives an identical bias current but no signal, as shown in Fig. 6. In addition, appropriately sized DC bias currents are added at other points in the circuit so that *all* integrators effectively have the same DC bias current and integrate the same offset. The output of the front end is then pseudo-differential: one of the signal ranges minus the REF range. To first order, this configuration cancels any bias-related offsets and extracts only the signal from all the ranges. An additional analog storage capacitor pipeline is added to the back end of the pixel for the REF range, so that the back end can acquire and read out in pseudo-differential fashion.

### 3.2 First Prototype Results

A first prototype chip was produced with some single-pixel test structures and with a 32 x 32 array of pixels. Along with front end circuitry, a rudimentary back end was also included in each pixel, consisting of several analog pipeline storage caps for each range, including the REF range. Pseudo-differential readout is then performed by simultaneously placing a storage capacitor from the desired range into a signal readout amplifier and a storage capacitor from the REF range into a reference readout amplifier.

Fig. 7 shows results of measurements taken when a wide range of input signal amplitudes is injected to a pixel in the array and a pseudo-differential readout is performed for all three ranges. The charge gain of range HG is measured to be 8.8 mV/fC, close to the design value, and the range scaling factors are close to 32, as expected. A dynamic range of $10^5$ is demonstrated. For this measurement, the readout amplifier biases were not optimized, which manifests as a back-end readout linear range of somewhat less than 500 mV. However, in the single pixel test structure, all three ranges are shown to have a linear range (error < 1%) of well over 500 mV.

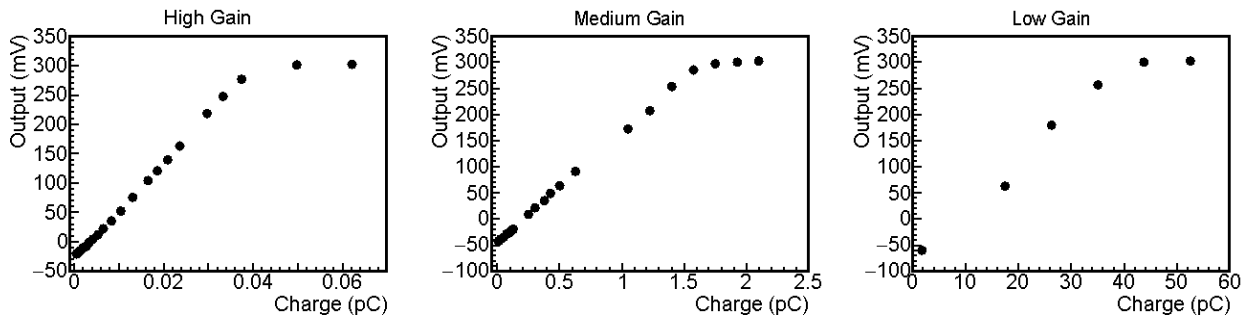

Fig. 7. First prototype multi-range response measurements.

Noise measurements were made on the single pixel test structure. The front-end noise is found to have three significant components: (1) the kTC noise of the integrator reset, (2) the voltage noise of the integrator voltage buffer, (3) and the integrated shot noise of the DC bias current. These noise components are present in both the signal and reference integrators. Table 1 shows both calculated and measured values for these noise components on the HG range for a pseudo-differential readout. For a single-ended readout of just the signal (and not the REF range), the noise would be smaller by $\sqrt{2}$. The total measured noise of 810 $e^-$ is higher than desired for measuring one photon (2000 $e^-$). Therefore, although the current splitter approach is shown to enable integration of large signals in a small pixel, the noise must be reduced significantly in order to attain the full dynamic range.

| Noise source | Calculated | Measured |
|---|---|---|
| Shot Noise (50 ns integration) | 612 $e^-$ | 720 $e^-$ |
| kTC Noise, $C_{int}$ = 160 fF | 280 μV → 199 $e^-$ | 297 μV → 211 $e^-$ |
| Voltage Buffer Noise | NA | 427 μV → 303 $e^-$ |
| Total (measured) | | 810 $e^-$ |

Table 1. First prototype range high-gain noise components

### 3.3 Second Prototype Design

A second prototype was designed with the goal of optimizing performance and significantly reducing the noise. Two strategies were employed: (1) a significant reduction in integrator capacitance value, which both reduces the kTC noise and increases the charge gain (thereby reducing the input-referred voltage buffer noise), and (2) a significant reduction in DC bias current, made possible by optimizing the range scaling.

With an open-loop (passive) integrator, the effective integration capacitance is the sum of the explicit physical capacitor and all the parasitic capacitors on that node. Since the parasitic capacitors on the integrator node are mostly diode junctions from MOS and NPN transistors, their value changes as a function of voltage. This affects the integrator linearity and therefore limits how small the total integrator capacitance value can be. However, with attention to transistor sizing, the variation of parasitic capacitance can be minimized. See Fig. 8, which shows that the parasitic capacitors consist of both P–N and N–P diode junctions. As the voltage changes, one junction polarity will be increasing in capacitance and the other will be decreasing. Proper sizing of the devices allows some degree of cancellation of the nonlinearities. In this second prototype, the explicit integrator capacitors were lowered to 45 fF for a total integration capacitance of about 60 fF. Optimization of the junction sizes was performed to maintain adequate linearity.

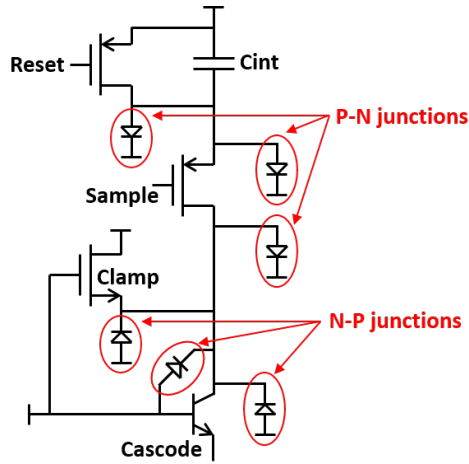

Fig. 8. Parasitic capacitances of the passive integrator.

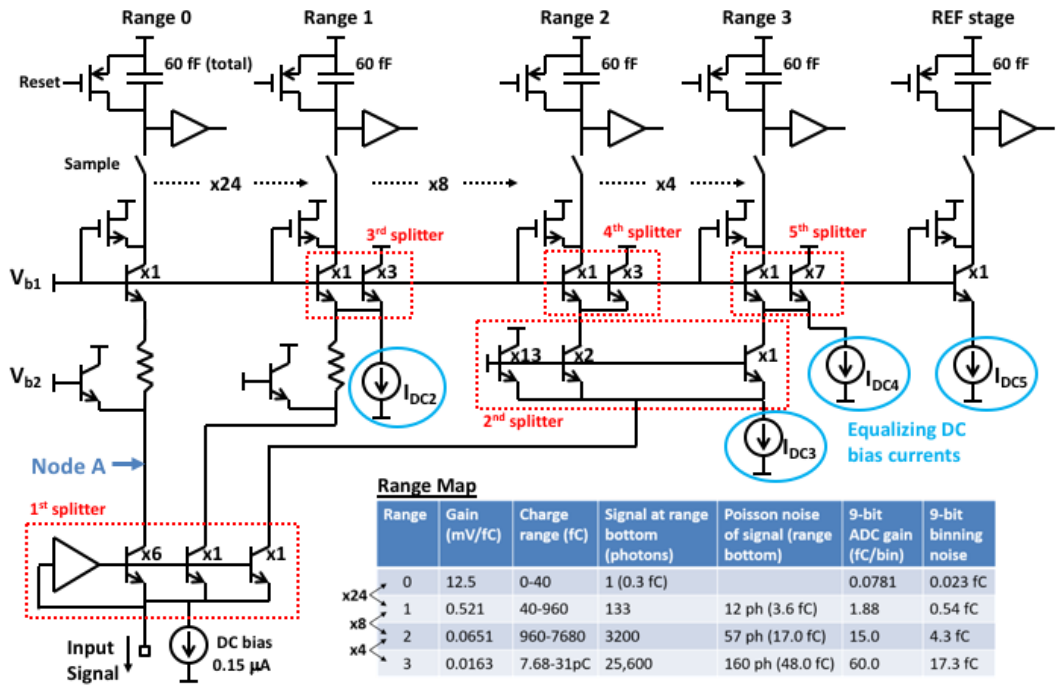

Fig. 9. Simplified schematic of the second front-end prototype.

To reduce the shot noise contribution, the DC bias current must be significantly lowered (by a factor of four for a factor of two in shot noise reduction). The only way to reduce the bias current while maintaining small-signal frequency response is to significantly reduce the capacitance on Node A. In the first prototype, this capacitance was dominated by the 32 collector junctions of the first range of the input current splitter. Fig. 9 is a simplified schematic of the second front end prototype with more optimal range scaling. By reducing the scale factor in the input current splitter, adding more downstream current splitters, and adding one more range (which allows smaller scale factors between ranges), the number of collectors on Node A is reduced from 32 to 6 while preserving overall dynamic range. The REF range is retained, and all integrators are identical. As in the first prototype, DC current sources are added at various nodes in the circuit so that all integrators receive the same DC bias current. The range map of this configuration in the lower right of Fig. 9 shows an additional benefit of a reduction in the number of ADC bits required, from 10 to 9. Since the charge gain on Range 0 has been increased, even with the loss of an ADC bit the charge binning noise at the bottom is not too different from the first prototype. By drastically reducing the number of input current splitter NPN transistors, the required DC bias current for adequate frequency response is reduced by a factor of four to about 0.15 µA.

### 3.4 Second Prototype Results

This second front-end prototype was produced only as a set of single pixel test structures, not as an array of pixels. The Range 0 measured charge gain is indeed higher at 12.6 mV/fC, which translates to an integrator capacitance of 60 fF (45 fF plus parasitics). The integrators are still linear (< 1% error). In a test structure of just the input current splitter, the measured split ratios are close to expected, and those split ratios stay constant (to within 0.1%) over all input currents up to 9 mA. As intended, adequate frequency response is attained with a lower bias current of 0.15 µA. Additionally, as shown in Table 2, the noise on Range 0 is reduced by about a factor of two as expected.

| Noise source | Calculated | Measured |
| --- | --- | --- |
| Shot Noise (50 ns integration) | 306 e$^-$ | 331 e$^-$ |
| kTC Noise, $C_{int}$ = 160 fF | 374 µV → 185 e$^-$ | 364 µV → 180 e$^-$ |
| Voltage Buffer Noise | NA | 354 µV → 175 e$^-$ |
| Total (measured) | | 415 e$^-$ |

Table 2. Second prototype range 0 noise components

### 3.5 Summary and Proposed Third Prototype

Through the process of designing and measuring the first two front end prototypes, several important points emerged: (1) the current splitting approach is effective in offering unprecedented dynamic range in a small pixel, (2) there is some design freedom in the number and arrangement of ranges and splitters that can be used in order to optimize range scaling and performance for a given application, (3) the noise performance can be improved by lowering the DC input bias current, and (4) the minimum bias current depends on the frequency response (and therefore on the parasitic capacitance) at the critical Node A.

After measuring the second prototype, a concentrated effort to lower the capacitance on Node A to the bare minimum was undertaken, resulting in the schematic of Fig. 10, which has been simulated and is proposed as a third prototype. The principal strategy in this circuit is to reduce the input DC bias current enough that the pseudo-differential configuration is no longer necessary, thereby significantly reducing the noise by both lowering the bias current and eliminating the REF stage. The Node A parasitic capacitance is again significantly reduced, and only enough DC bias current is applied to all the other ranges to insure the minimum required bandwidth on each range. The offsets due to integrated bias current on Ranges 1-3 are now small enough that they are not significant, and the reduced but still possibly problematic offset on Range 0 can be canceled with an optional $I_{comp}$ current if necessary (at the cost of less reduction of noise on that range).

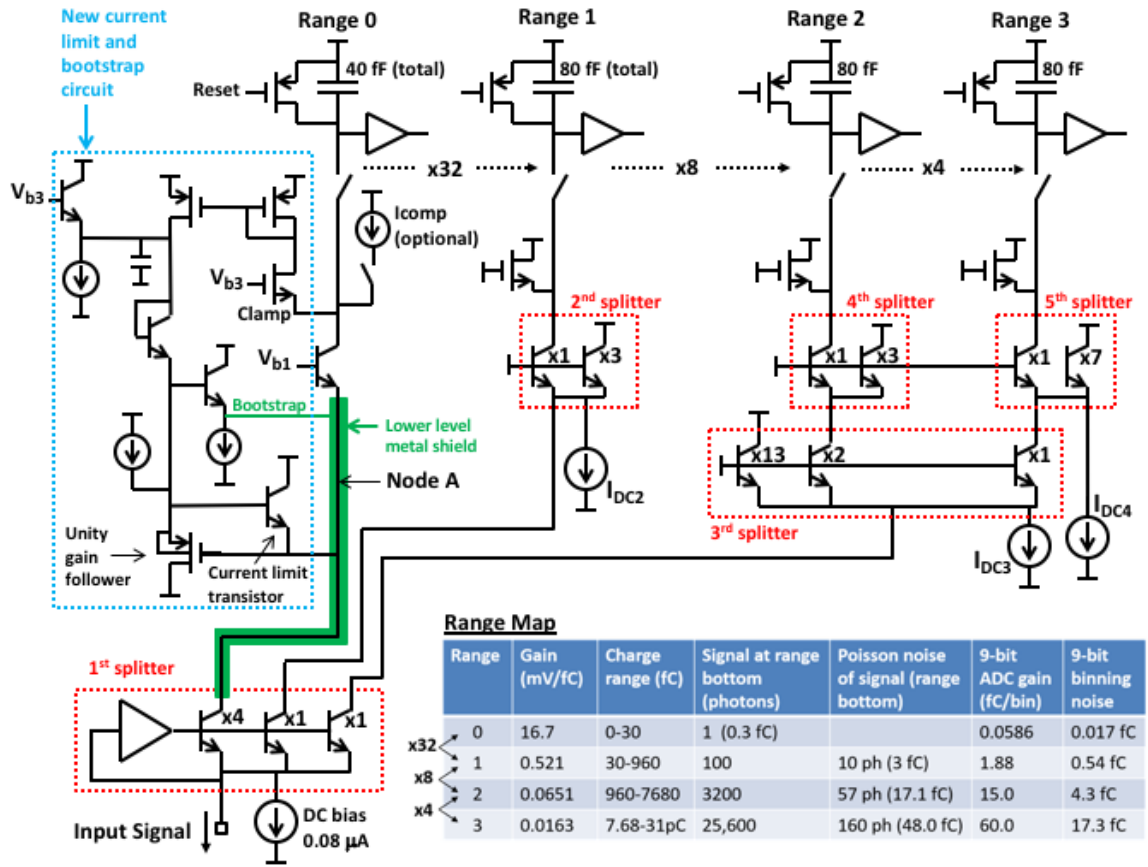

Fig. 10. Simplified schematic of the proposed third front-end prototype.

    To further reduce the number of collectors on Node A, the number of input splitter transistors on Range 0 is reduced again, from six to four, by now doing some of the range scaling with the integrator capacitance value. Since these capacitors are very small, this has little impact on the area required. The new range map is shown in Fig. 10. Also, a new technique is devised for the current limit function that effectively removes its capacitance for small signals. This is accomplished by bootstrapping the current limit NPN base to Node A with a unity gain follower, so that its base follows its emitter. Another bootstrap buffer is driven from the current limit NPN base node, so that this buffer's low impedance output can drive a metal shield that is placed under the metal wiring net of Node A. This effectively removes the significant parasitic wiring capacitance to the substrate of Node A. For large signals, the current limit function is initiated not by a series resistor as before, but by the integrator clamp transistor. When the Range 0 integrator goes beyond its valid range, it turns on this clamp transistor. The clamp current is mirrored and pulls up the base of the current limit transistor significantly, which promptly turns it on and diverts the excess Node A current to a virtual ground.

    In summary, by reducing the number of collectors on Node A to a bare minimum and effectively removing the capacitance of the wiring and the current limit transistor through bootstrapping, the Node A capacitance is reduced to its bare minimum. This allows a reduction of the DC input bias current to approximately 0.08 µA, giving a $\sqrt{2}$ improvement in the noise. Removal of the REF range yields another $\sqrt{2}$ noise improvement (unless the optional $I_{comp}$ current is needed), with a resultant front-end equivalent input noise charge of around 200 $e^-$. This is ten times below the one photon signal level, giving an adequate signal to noise ratio.

## 4. Conclusion

    Through two prototype runs, the current splitting technique has been demonstrated to significantly increase the maximum charge that can be integrated in a small pixel. The second prototype successfully lowered the

equivalent input charge noise to a level close to what is required for single-photon detection. A third prototype is proposed that should further reduce the noise, resulting in a full dynamic range of $10^5$.

**Acknowledgments**

The author would like to acknowledge Davide Braga for doing the back end design and helping with implementation of the test protocol, and to Lou DalMonte for help with the physical test setup.

This material is based upon work supported by the U.S. Department of Energy, Office of Science, Office of Basic Energy Sciences, under Award Number: DE-SC0017170.

This manuscript has been authored by Fermi Research Alliance, LLC under Contract No. DE-AC02-07CH11359 with the U.S. Department of Energy, Office of Science, Office of High Energy Physics.

The work used resources of the Advanced Photon Source, a U.S. Department of Energy (DOE) Office of Science User Facility operated for the DOE Office of Science by Argonne National Laboratory under Contract No. DE-AC02-06CH11357.